\documentclass[prx,twocolumn,floatfix]{revtex4-2}
\usepackage{graphicx}  
\usepackage{dcolumn}   
\usepackage{comment}
\usepackage{bm}        
\usepackage{amssymb}   
\usepackage{amsmath}
\usepackage{mathrsfs}
\usepackage{epigraph}
\usepackage{gensymb}
\usepackage{lmodern}
\usepackage{lipsum}
\usepackage{marvosym}
\usepackage{tipa}
\usepackage{bbold}
\usepackage{dsfont}
\usepackage{esint}
\usepackage{xcolor}
\usepackage{braket}
\usepackage{appendix}
\usepackage{natbib}
\usepackage{bbold}
\usepackage[colorlinks=true, linkcolor=blue, urlcolor=blue,
citecolor=blue]{hyperref}

\begin{document}
	
	\title{Quantum-classical approach to spin and charge pumping and the ensuing radiation in THz spintronics: Example of ultrafast-light-driven  Weyl antiferromagnet Mn$_3$Sn}
	
	% [<] Authors
	\author{Abhin Suresh}
	\affiliation{Department of Physics and Astronomy, University of Delaware, Newark DE 19716, USA}
	\author{Branislav K.~Nikoli\'c}
	\email{bnikolic@udel.edu}
	\affiliation{Department of Physics and Astronomy, University of Delaware, Newark DE 19716, USA}
	
	\begin{abstract}
		The interaction of fs laser pulse with magnetic materials has been intensely studied for more than two decades in order to understand ultrafast demagnetization in single magnetic layers or THz emission from their bilayers with nonmagnetic spin-orbit (SO) materials. However, in contrast to well-understood spin and charge pumping by dynamical magnetization in spintronic systems driven by microwaves or current injection, analogous processes in light-driven magnets and radiation emitted by them remain largely unexplained due to {\em multiscale} nature of the problem. Here we develop a multiscale quantum-classical formalism---where conduction electrons are described by quantum master equation (QME) of the Lindblad type; classical dynamics of local magnetization is described by the  Landau-Lifshitz-Gilbert (LLG) equation; and incoming light is described by classical vector potential while outgoing electromagnetic radiation is computed using the Jefimenko equations for retarded electric and magnetic fields---and apply it to a bilayer  of antiferromagnetic Weyl semimetal Mn$_3$Sn, hosting noncollinear local magnetization, and SO-coupled nonmagnetic material. Our QME+LLG+Jefimenko scheme makes it possible to understand how fs laser pulse generates directly spin and charge pumping and electromagnetic radiation by  Mn$_3$Sn layer, including {\em both} odd and even high harmonics (of the pulse center frequency) up to order $n \le 7$. The directly pumped spin current then exerts spin torque on local magnetization whose dynamics, in turn, pumps additional spin and charge currents radiating in the THz range. By switching on and off LLG dynamics and SO couplings, we unravel which microscopic mechanism contribute the most to emitted THz radiation---{\em charge pumping by local magnetization of Mn$_3$Sn in the presence of its own SO coupling} is far more important than standardly assumed (for other types of magnetic layers) spin pumping and subsequent spin-to-charge conversion within the adjacent nonmagnetic SO-coupled material. 
	\end{abstract}

	\maketitle
	
	\section{Introduction}\label{sec:intro}
	
\begin{figure}
		\centering
		\includegraphics[width = \linewidth]{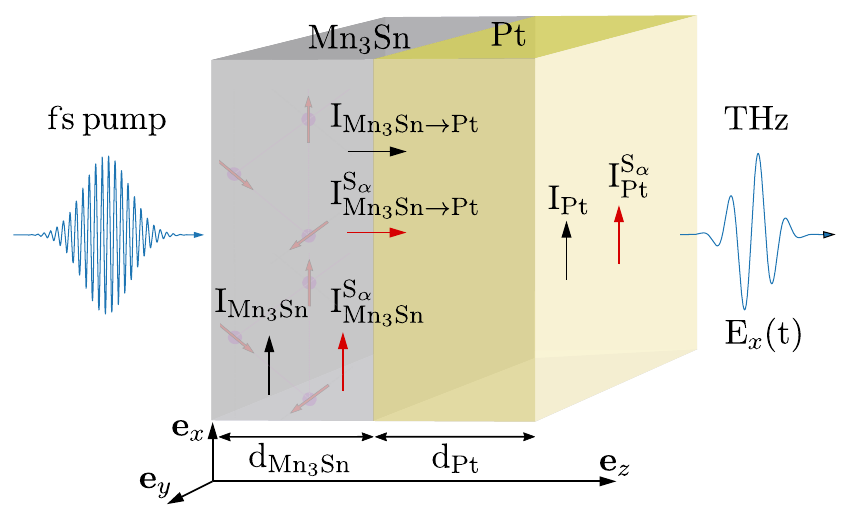}
		\caption{Schematic view of a bilayer comprised of: a layer of noncollinear Weyl AF  Mn$_3$Sn of thickness $d_{\mathrm{Mn}_3\mathrm{Sn}}=2$ MLs; and nonmagnetic SO-material (such as Pt) of thickness $d_\mathrm{Pt}=1$ ML. The magnetic layer is modeled by quantum TB [Eq.~\eqref{eq:Htot}] while its LMMs are described by  classical Heisenberg [Eq.~\eqref{eq:Hcls}] Hamiltonians, and the second layer is modeled by quantum TB Hamiltonian including the Rashba SOC.  Both MLs of Mn$_3$Sn are assumed to be irradiated by fs laser pulse of 800 nm center wavelength, as illustrated on the left. The  dynamics of its LMMs is animated in movies {\tt during\_laser\_pulse.mp4} and {\tt after\_laser\_pulse.mp4} in the SM~\cite{sm}. The pulse and/or the  dynamics of LMMs generate interlayer ($I_{\mathrm{Mn}_3\mathrm{Sn} \rightarrow \mathrm{Pt}}$ and $I^{S_\alpha}_{\mathrm{Mn}_3\mathrm{Sn} \rightarrow \mathrm{Pt}}$) and intralayer ($I_{\mathrm{Mn}_3\mathrm{Sn}}$, $I_{\mathrm{Mn}_3\mathrm{Sn}}^{S_\alpha}$, $I_\mathrm{Pt}$ and $I_\mathrm{Pt}^{S_\alpha}$)  charge ($I$) and spin ($I^{S_\alpha}$) currents. The time-dependent charge currents and densities determine, via the Jefimenko~\cite{Jefimenko1966,Ridley2021} equations [Eqs.~\eqref{eq:efield} and ~\ref{eq:bfield}], emitted THz radiation illustrated on the right, as well as high harmonics of the center frequency $\Omega_0$ of the laser pulse in the emitted EM radiation in higher than THz range.}
		\label{fig:fig1}
	\end{figure}

	Laser-induced ultrafast demagnetization~\cite{Beaurepaire1996,Kirilyuk2010} and the ensuing THz spin transport and electromagnetic (EM) radiation---from a single magnetic layer~\cite{Rouzegar2022}, or in contact~\cite{Rouzegar2022,Seifert2022,Bull2021,Wu2021,Seifert2016,Wu2016,Chen2018} with an additional layer [Fig.~\ref{fig:fig1}] of a nonmagnetic (NM) material hosting strong spin-orbit coupling (SOC) effects---are central phenomena in femtomagnetism and THz spintronics. Besides fundamental interest in ultrafast coupled charge and spin dynamics in systems where moving hot electrons interact with localized magnetic moments (LMMs)~\cite{Gillmeister2020,Wang2017a}, {\em far-from-equilibrium magnets} are also of great interest for spintronics applications. For example, magnetization of a single layer of such material can be reversed~\cite{Kimel2019} on sub-ps time scale for digital memory applications, in contrast to standard current-driven switching via spin torque in near-equilibrium magnets which takes much longer $\sim 100$ ps time. In addition, driving bilayers by low-cost and low-power femtosecond laser---where ferromagnetic metal (FM), or metallic or insulating antiferromagnet (AF), is attached to a nonmagnetic SO-material---have opened new avenues~\cite{Seifert2022,Seifert2016,Wu2016,Chen2018} for highly efficient table-top emitters of ultrabroad band 1--30 THz EM radiation. Conversely,  other THz solid-state emitters, such as standard ZnTe crystal, rely solely on physics related to electron charge and deliver emission spectra with substantial gaps~\cite{Seifert2016} and with low peak intensity~\cite{Wu2016} even for \mbox{$\sim 100$ $\mu$m} thick crystals. The usage of magnetic multilayers also makes it possible to tailor properties of spintronics THz emitters within desired frequency range, such as to enhance~\cite{Chen2018} THz signal at the lower THz frequency range (0.1--0.5 THz).
	
	Despite more than 20 years of intense studies~\cite{Beaurepaire1996}, the underlying physics of the laser-induced demagnetization is still under scrutiny with one of the main puzzles being different paths for angular momentum transfer~\cite{Chen2019a} and their interplay to produce ultrafast time evolution of magnetization. In addition, it is widely believed that laser-induced magnetization dynamics generates primarily~\cite{Malinowski2018} spin current~\cite{Seifert2022,Seifert2016,Wu2016}. The spin current then has to be injected~\cite{Lu2021} into the adjacent nonmagnetic layer to be efficiently converted into transient charge current as the source of THz radiation. The experimentally explored mechanisms of spin-to-charge conversion include the inverse spin Hall effect~\cite{Seifert2022,Seifert2016,Wu2016} from the bulk SOC of nonmagnetic layer; interfacial SOC~\cite{Jungfleisch2018a}; and interfacial skew-scattering~\cite{Gueckstock2021}. 
	
		%----------------------------------------------------------
	%----------------------------------------------------------
	
	While  pumped (term used to signify current generation in the absence of any bias voltage~\cite{Vavilov2001,FoaTorres2005,Tserkovnyak2005,Bajpai2019}) spin currents by laser-driven magnetic layer, with short attenuation length \mbox{$\sim 10$ nm}~\cite{Gorchon2022}, have been observed experimentally~\cite{Eschenlohr2017,Gorchon2022},  microscopic mechanisms behind them and their relation (or even necessity~\cite{Malinowski2018}) to demagnetization remain heavily debated~\cite{Malinowski2018,Lichtenberg2022}. For example,  optically excited hot electrons~\cite{Malinowski2018} become spin-polarized by magnetic layer to comprise spin current in the so-called ``superdiffusive mechanism''~\cite{Battiato2010,Battiato2012,Malinowski2018,Gupta2022}, whose flowing out of the FM layer then contributes to demagnetization. Conversely, the cause and effect are reversed  in the so-called ``$dM/dt$ mechanism''~\cite{Koopmans2010,Lichtenberg2022,Tveten2015} where demagnetization is due to excitation of magnons which then transfer lost angular momentum to spin-polarized conduction electrons, so that optically excited hot electrons~\cite{Malinowski2018} do not play a major role in such generation of spin current proportional to the time derivative of the magnetization. These mechanisms are also expected to leave different imprints on the temporal profile of generated spin currents vs. the laser-pulse intensity, as probed in  recent experiments~\cite{Lichtenberg2022}. 
	
	Also, the notion that spin-to-charge conversion is necessary for THz radiation neglects possibility of charge current pumping directly by magnetization dynamics when proper symmetries are broken~\cite{Chen2009,Bajpai2019} or SOC is present~\cite{Mahfouzi2012,Ciccarelli2015} in the bulk or at the interface 
	of magnetic layer. For instance, both types of SOC [such as, the fourth term on the right hand side (RHS) of Eq.~\eqref{eq:H1} and the second term on the RHS of Eq.~\eqref{eq:HPt}] are operative in magnetic bilayer [Fig.~\ref{fig:fig1}] of Mn$_3$Sn chosen as an illustrative example in this study. Note that time-dependent magnetization of a single FM layer itself emits THz radiation of magnetic-dipole type, but its intensity is orders of magnitude smaller~\cite{Rouzegar2022} than the signal generated from magnetic bilayers with properly pumped and/or enhanced time-dependent charge current as the radiation source (Sec.~\ref{sec:jefimenko}).
	
	%----------------------------------------------------------
\begin{figure}
	\centering
	\includegraphics[width = \linewidth]{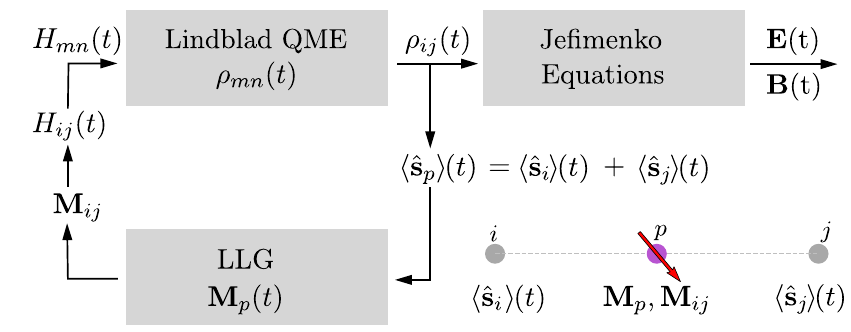}
	\caption{Self-consistent scheme of QME+LLG+Jefimenko quantum-classical approach, where time-dependent density operator [$\hat{\rho}(t)$ and its matrix representation ${\bm \rho}(t)$] of nonequilibrium electrons is obtained by solving the Lindblad QME. The density operator determines the expectation value of spin and charge currents and densities, where nonequilibrium spin density  $\langle \hat{\mathbf{s}}_p \rangle(t)$ generates spin torque $\propto \langle \hat{\mathbf{s}}_p \rangle(t) \times  \mathbf{M}_p(t)$ in the LLG equation for classical dynamics of LMMs $\mathbf{M}_p(t)$ of a magnetic material whose time-dependence, in turn, adds time-dependent term (in addition to time-dependence due to the laser pulse) into the quantum Hamiltonian of electrons. Finally, thus computed time-dependent charge currents and densities are fed as sources into the Jefimenko equations~\cite{Jefimenko1966,Ridley2021} for retarded electric $\mathbf{E}(t)$ and magnetic $\mathbf{B}(t)$ fields of emitted EM radiation. Inset in the lower right corner illustrates how both nonequilibrium spin densities at Sn sites $i$ and $j$ contribute [Eq.~\eqref{eq:noneqspinp}] to $\langle \hat{\mathbf{s}}_p \rangle(t)$ which interacts with LMM at site $p$ hosting magnetic atom of Mn.}
	\label{fig:fig2}
\end{figure}
%----------------------------------------------------------

	A popular, but qualitative, picture of complex processes ignited in laser-driven magnetic materials is offered by the so-called three-temperature model~\cite{Beaurepaire1996,Malinowski2018}---electrons, phonons and LMMs are assumed to be at three different temperatures and either optically excited hot electrons interact directly with LMMs or thermalized electrons interact with LMMs via secondary processes~\cite{Stiehl2022}. The quantitative modeling of these complex processes has been pursued via three major routes: ({\em i}) phenomenological and purely classical micromagnetic modeling~\cite{Ulrichs2018} of the dynamics of LMMs via the Landau-Lifshitz-Gilbert (LLG) equation~\cite{Evans2014,Ritzmann2020}, which excludes electrons and thereby does not allow us to understand how angular momentum flows between different subsystems~\cite{Hennecke2019,Tauchert2022} or how electrons comprise pumped spin and charge currents; ({\em ii}) phenomenological semiclassical modeling of laser excited hot electron transport, such as using the Boltzmann~\cite{Nenno2018,Nenno2019},  Vlasov~\cite{Hurst2018} or superdiffusive~\cite{Battiato2010,Battiato2012} transport equations including possible combination~\cite{Ritzmann2020} with atomistic  spin dynamics (i.e., the LLG equation) of LMMs; and ({\em iii}) microscopic  (i.e., Hamiltonian-based) quantum modeling via quantum master equations (QMEs)~\cite{Topler2021}, small-cluster exact time-propagation~\cite{Tows2019} or time-dependent density functional theory (TDDFT)~\cite{Krieger2015,Krieger2017,Dewhurst2018,Chen2019a}  employing single-particle tight-binding (TB)~\cite{Topler2021}, many-body second-quantized~\cite{Tows2019} or first-principles Hamiltonians, respectively.

	For example, quantum many-body~\cite{Tows2019} and TDDFT~\cite{Krieger2015,Krieger2017,Dewhurst2018} modeling highlight SOC induced spin-flips in both thin films and bulk or initial spin disorder~\cite{Chen2019a} as the major effects on demagnetization. Both TDDFT~\cite{Krieger2017} and QME approaches~\cite{Topler2021} find further reduction of magnetization in the course of its time evolution in thin films due to processes at interfaces  and surfaces. Microscopic quantum modeling also provides detailed insight into the flow of angular momentum~\cite{Chen2019a,Dewhurst2021,Hennecke2019,Tauchert2022} between electron spin, electron orbital degrees of freedom, and ionic lattice (see, e.g., Fig.~2 in Ref.~\cite{Chen2019a}). Furthermore, by switching on and off different terms in the Hamiltonian, one can pinpoint importance and magnitude of different mechanisms, such as that: direct coupling of light to spins (via the Zeeman term containing magnetic field of light) is irrelevant as spin motion cannot follow rapidly oscillating magnetic field of laser pulse~\cite{Chen2019a}; instead, it is light-orbital interaction which transfers energy and angular momentum to electrons while affecting their spins via SOC; the SOC-driven effects are ineffective in the first \mbox{$\simeq 10$ fs} of demagnetization process (see, e.g., Fig.~3 in Ref.~\cite{Siegrist2019}). 
	
	However, the semiclassical equations for electron transport {\em cannot} capture pumping of spin~\cite{Tserkovnyak2005} and charge currents~\cite{Chen2009,Ciccarelli2015,Mahfouzi2012} by {\em time-dependent quantum} system of electrons whose time-evolution is triggered both by laser pulse~\cite{Bajpai2019}  and  
	dynamical LMMs~\cite{Tserkovnyak2005}. 	Furthermore, the presence of noncollinear textures of LMMs, either intrinsically as in  Weyl AF Mn$_3$Sn or due to thermal fluctuations~\cite{Ghosh2022}, opens additional channels~\cite{Zhang2009b,Kim2012b,Yamane2011} for spin and charge pumping by quantum transport of conduction electrons surrounding~\cite{Petrovic2018,Bajpai2019a,Bajpai2020,Petrovic2021} such dynamical magnetic textures. For example, the  approach of Ref.~\cite{Hurst2018}, combining the Vlasov equation for electrons with the LLG equation~\cite{Evans2014,Ritzmann2020} for LMMs, requires three different ingredients (magnetic ground state, metal-vacuum interfaces near which the spin-up and spin-down dipoles are progressively dephased; and self-consistent electric field which mediates momentum transfers between the two types of dipoles at such interfaces) in order to obtain nonzero spin current by oscillating magnetic dipole due to purely electric excitation by laser pulse. We note that superdiffusive~\cite{Battiato2010,Battiato2012} or the Boltzmann equation approaches~\cite{Nenno2018,Nenno2019} do  model spin currents mediated by charge dynamics (as the difference between spin-up and spin-down charge currents), but they cannot capture light- or LMM-dynamics-induced pumping of spin and charge currents as purely 
	quantum time-dependent phenomenon.
	
	Although microscopic quantum approaches can in principle capture all currents generated by time-dependence of a quantum system of electrons (or electrons and LMMs if the latter are also treated quantum-mechanically~\cite{Tows2019}),  computing pumped currents has been bypassed in quantum many-body~\cite{Tows2019}  and TDDFT studies~\cite{Krieger2015,Krieger2017,Dewhurst2018,Dewhurst2021,Chen2019a} due to their primary focus on nonequilibrium magnetization vs. time. In QME approaches~\cite{Topler2021} different mechanisms of spin and charge current pumping, such as due to ultrafast laser pulse vs. due to subsequent dynamics of 
	much slower LMM, as illustrated by the movies in the Supplemental Material (SM)~\cite{sm}, were also not resolved.  
	
	Besides the need to calculate spin and charge currents, as well as  spin(current)-to-charge(current) conversion~\cite{Seifert2022,Seifert2016,Wu2016,Jungfleisch2018a,Gueckstock2021}, complete description of THz spintronics experiments also requires to compute how these currents emit THz radiation. This calculation is presently lacking from both QME and TDDFT approaches due to the need to couple their respective equations to the Maxwell equations in {\em multiscale} fashion. For example,  such a task has only very recently been initiated in TDDFT software development~\cite{Noda2019,Tancogne-Dejean2020}.
	
	In this study, we develop a {\em multiscale  quantum-classical} approach [Fig.~\ref{fig:fig2}]  where: quantum transport of electrons, including dissipation and decoherence effects due to coupling to an external bosonic bath as provided by phonons~\cite{Sold2018}, is described by QME~\cite{Topler2021} of the Lindblad type~\cite{Lindblad1976,Manzano2020} for electronic nonequilibrium one-particle density matrix; dynamics of classical LMMs is described via the LLG equation within the general framework of atomistic spin dynamics~\cite{Evans2014,Ritzmann2020}; and incoming light is described by classical vector potential, which couples to electrons in QME, while properly retarded electric and magnetic fields of emitted EM radiation are computed from the Jefimenko equations~\cite{Jefimenko1966}. The Jefimenko equations are also adapted~\cite{Ridley2021} to employ time-dependent bond charge currents~\cite{Nikolic2006,Ridley2021} and charge densities as field sources extracted from QME-based quantum transport treatment of electrons. We apply these QME+LLG (for equilibration) and QME+LLG+Jefimenko (for laser pulse driven nonequilibrium dynamics and emitted EM radiation) calculations to a bilayer  [Fig.~\ref{fig:fig1}] of noncollinear Weyl AF Mn$_3$Sn~\cite{Yang2017} in contact with SO-coupled material like Pt, where the bilayer is modeled on the TB lattice by widely used quantum Hamiltonian of Weyl conduction electrons within Mn$_3$Sn~\cite{Liu2017} and classical Hamiltonian of its noncollinear LMMs~\cite{Nomoto2020}. 
	
	Besides having well-defined quantum and classical Hamiltonians, as the only input required in our microscopic approach, the choice of Mn$_3$Sn as an illustrative example is also motivated by recent experiments~\cite{Zhou2019b} observing THz emission from bilayers of  Weyl AF Mn$_3$Sn and SO-materials like Pt. At first sight, these observations, as well as others~\cite{Qiu2020,Rongione2022} involving AF spintronics THz emitters,  are counter-intuitive as AF materials exhibit no net magnetization while the principal magneto-optical effects are linear in the net magnetization. Thus, application of our microscopic quantum-classical approach to AF spintronic THz emitters, treating coupled quantum dynamics of conduction electrons and classical atomistic spin dynamics of  LMMs centered 
	on individual magnetic atoms of AF crystalline lattice, allows one to resolve a number of such puzzles. 	
	
	The paper is organized as follows. Section~\ref{sec:models} explains our QME and LLG methodologies with their respective input in the form of quantum and classical Hamiltonians. In the same Section, we also explain how quantum-transport-computed charge currents and densities are transferred into the Jefimenko equations to obtain retarded electric and magnetic fields of emitted EM radiation. We discuss results for pumped spin and charge currents in Sec.~\ref{sec:currents} and for emitted EM radiation in Sec.~\ref{sec:emittedthz}. In addition, since the analysis of the energy flux of incoming and outgoing electromagnetic radiation, and thereby defined efficiency of their conversion into each other, is lacking in THz spintronics literature, Sec.~\ref{sec:poynting} shows computation of their respective Poynting vectors~\cite{Poynting1884}, their ratios, as well as angular dependence of  Poynting vector for outgoing radiation. We conclude in Sec.~\ref{sec:conclusions}. We also provide three movies as the SM~\cite{sm}, which animate: equilibration of LMMs within Mn$_3$Sn layer and their disordering [with respect to configuration shown in Fig.~\ref{fig:fig1}] after interaction with room temperature electrons is switched on; dynamics of LLMs and spin and charge current pumping during the application of fs laser pulse driving electrons out of equilibrium; and dynamics of LLMs and spin and charge current pumping after the laser pulse has ceased. 
	
	\section{Model and methods}\label{sec:models}
	
	\subsection{Quantum and classical Hamiltonians}\label{sec:hamiltonians}
	
	The subsystem of conduction electrons within the magnetic bilayer in Fig.~\ref{fig:fig1} is described by a quantum Hamiltonian which we split into three contributions
	\begin{equation}\label{eq:Htot}
		\hat{H}_{\rm tot}(t) = \hat{H}_{\rm Mn_3Sn}(t) + \hat{H}_{\rm Pt} + \hat{H}_{\rm Mn_3Sn-Pt}. 
	\end{equation}
	Here $\hat{H}_{\rm Mn_3Sn}(t)$ describes Weyl electrons within Mn$_3$Sn layer 
	\begin{equation}\label{eq:H1}
		\begin{split}
			\hat{H}_{\rm Mn_3Sn}(t) =\ & \sum_{\left<i,j\right>_{xy}} \gamma^{xy}_{ij}(t)
			\hat{\mathbf{c}}^{\dagger}_{i}\hat{\mathbf{c}}_{j} +\ \sum_{\left<i,j\right>_{z}} \gamma^z_{ij}(t) \hat{\mathbf{c}}^{\dagger}_{i}\hat{\mathbf{c}}_{j}\\ 
			-\ &J_{sd}\sum_{\left<i,j\right>_{xy}}\hat{\mathbf{c}}^{\dagger}_{i} \hat{\boldsymbol{\sigma}}\hat{\mathbf{c}}_{j} \cdot \mathbf{M}_{ij}(t)\\
			+\ &i\lambda_z\sum_{\left<i,j\right>_{xy}}(-1)^{\xi_{i,j}}\hat{\mathbf{c}}^{\dagger}_{i} \hat{\sigma}_z\hat{\mathbf{c}}_{j},
		\end{split}
	\end{equation}
	and it is chosen as a minimal TB model~\cite{Liu2017}, where  $\langle i,j \rangle$ signifies that only the nearest-neighbor (NN) hoppings are taken into account, with single orbital centered at each Sn site $i$. Those sites $i$ form a hexagonal lattice of lattice constant \mbox{$a_0$}. The time-dependence of $\hat{H}_{\rm Mn_3Sn}(t)$ stems from both the laser pulse [first two terms on the RHS of Eq.~\eqref{eq:H1}] and dynamics of LMMs [third term on the RHS of Eq.~\eqref{eq:H1}]. The second contribution in Eq.~\eqref{eq:Htot} models SO-coupled electrons within the nonmagnetic layer in Fig.~\ref{fig:fig1}
	\begin{equation}\label{eq:HPt}
		\begin{split}
			\hat{H}_{\rm Pt} = \gamma_{\rm Pt}\sum_{\langle i,j \rangle} \hat{\mathbf{c}}^{\dagger}_{i} \hat{\mathbf{c}}_j + \sum_{\langle i,j \rangle} \hat{\mathbf{c}}^\dagger_i {\bm \gamma}_{ij}^{\rm SOC} \hat{\mathbf{c}}_j,
		\end{split}
	\end{equation}
	where the Rashba SOC~\cite{Manchon2015} (in TB form~\cite{Nikolic2006}) is used in the second term on the RHS. This choice  inspired by first-principle calculations~\cite{Park2013} of band structure at interfaces of heavy metals like Pt and magnetic materials, or experiments in THz spintronics~\cite{Jungfleisch2018a} interpreted under the assumption of Rashba SO-coupled interface between magnetic and nonmagnetic layers.  
	The third term in Eq.~\eqref{eq:Htot}
	\begin{equation}\label{eq:Hinter}
		\hat{H}_{\rm Mn_3Sn-Pt} = \gamma_\mathrm{inter} \sum_{\langle ij \rangle} \hat{\bm c}^{\dagger}_{i} \hat{\bm c}_{j},
	\end{equation}
	describes hopping between sites $i$ of magnetic Mn$_3$Sn layer and sites $j$ of nonmagnetic Pt layer in Fig.~\ref{fig:fig1}. The notation employed in Eqs.~\eqref{eq:H1}, ~\eqref{eq:HPt} and ~\eqref{eq:Hinter} is:  \mbox{$\hat{\mathbf{c}}^{\dagger}_i = (\hat{c}^{\dagger}_{i\uparrow} \  \hat{c}^{\dagger}_{i\downarrow})$} is the row vector containing operators $\hat{c}^{\dagger}_{i,\sigma}$ that create electron of spin $\sigma$ at site $i$ and $\hat{\mathbf{c}}_i = \big( \hat{\mathbf{c}}^{\dagger}_i \big)^\dagger$ is the corresponding column vector of annihilation operators; \mbox{${\hat{\bm \sigma}} =(\hat{\sigma}_x,\hat{\sigma}_y,\hat{\sigma}_z)$} is the vector of the with Pauli matrices; $\mathbf{M}_{ij}$ is a unit vector of classical LMM at Mn site located in the middle of the bond between Sn sites $i$ and $j$~\cite{Liu2017,Nomoto2020}; $\gamma_{xy}(t)$ and $\gamma_z(t)$ are the NN hopping parameters~\cite{Liu2017} within monolayer (ML) of Mn$_3$Sn (i.e., within the $xy$-plane in the coordinate system of Fig.~\ref{fig:fig1}) and in between two such MLs  of Mn$_3$Sn, respectively, with their time dependence stemming from applied laser pulse (Sec.~\ref{sec:currents}); $\gamma_{\rm Pt}$ is NN hopping within ML of Pt; $\gamma_\mathrm{inter}$ is hopping between the last ML of Mn$_3$Sn layer and the first ML of Pt; and $J_{sd}$ is the strength of $sd$ exchange interaction~\cite{Cooper1967} between electron spin and classical LMMs.  Within ML of Mn$_3$Sn, SOC of strength $\lambda_z$ has sign which depends~\cite{Liu2017} on the chirality $\xi_{ij}$ of the bond ${\bf r}_{ij}$, such as  \mbox{$\xi_{ij}=+1$} for the NN bond of Sn atoms along the [100], [010], and $[\overline{1}\overline{1}0]$ directions, and \mbox{$\xi_{ij}=-1$} for the other three NN  bonds. We use \mbox{$\gamma^{xy}_{ij}(t=0)=\gamma=1$ eV}, \mbox{$\gamma_{\rm Pt}=\gamma_\mathrm{inter}=1$ eV}, \mbox{$\gamma^z_{ij}(t=0)=\gamma^z=0.5$ eV}, \mbox{$J_{sd}=0.5$ eV}, and \mbox{$\lambda_z=0.5$ eV}. The second term in Eq.~\eqref{eq:HPt} is the tight-binding~\cite{Nikolic2006}  version of the Rashba SOC~\cite{Manchon2015} where $2 \times 2$ matrix ${\bm \gamma}_{ij}^{\rm SOC}$ (in the electron spin space) contains hoppings  of magnitude  \mbox{$\gamma^\mathrm{SOC}=1$ eV}, as given in Eq.~(12) of Ref.~\cite{Nikolic2006}.

	%\begin{equation}\label{eq:rso}
	%    {\bm \gamma}_{ij}^{\rm RSO}=
	%    \begin{cases} 
		%    -i\gamma_{\rm RSO} \hat{\sigma}_y\ &(i = j + \mathbf{e}_\textit{x}) \\
		%    +i\gamma_{\rm RSO} \hat{\sigma}_x\  &(i = j + \mathbf{e}_\textit{y}).
		%    \end{cases}
	%\end{equation}
	%where we use \mbox{$\gamma_{\rm RSO} = 1$ eV} and 

	The classical LMMs at Mn sites $p$ are governed by their own extended classical Heisenberg Hamiltonian~\cite{Nomoto2020} 
	\begin{equation}\label{eq:Hcls}
		\begin{split}
			\mathcal{H}(t) =& J_1\sum_{\left <p,q\right >_{xy}}\mathbf{M}_p\cdot\mathbf{M}_q\ +\ J_2\sum_{\left <p,q\right >_z}\mathbf{M}_p\cdot\mathbf{M}_q\\
			&+\ \sum_{\left <p,q\right >_{xy}} \mathbf{D}_{pq}\cdot(\mathbf{M}_p \times \mathbf{M}_q) - K\sum_{p}(\hat{\bf n}_p\cdot \mathbf{M}_p)^2\\
			&- J_{sd} \sum_{p} \langle \hat{\bf s}_p \rangle \cdot \mathbf{M}_p (t).
		\end{split}
	\end{equation}
	Note that here we use notation $\mathbf{M}_{p} \equiv \mathbf{M}_{ij}$, while  $\mathbf{M}_{ij}$ is more clarifying notation for Eq.~\eqref{eq:H1}. In addition, $J_1$ and $J_2$ are isotropic exchange couplings between intra-ML and inter-ML NN LMMs; $\mathbf{D}_{pq}$ is the intra-ML Dzyaloshinskii-Moriya interaction specified by the vector \mbox{$\mathbf{D}_{pq}= D\hat{z} + D'\hat{z} \times \mathbf{e}_{\textit{pq}}$}, where $\bf{e}_{\textit{pq}}$ is the unit vector oriented from Mn site $p$ to Mn site $q$; $\hat{\mathbf{n}}_p$ is a unit vector characterizing the local easy axis at site $p$, oriented along the direction between the spin $p$ and either of its NN Sn ions in the plane. The values of parameters in Eq.~\eqref{eq:Hcls} satisfy \mbox{$J_1,J_2 \gg D$} and \mbox{$D' \gg K$}, where we use~\cite{Nomoto2020} \mbox{$J_1=J_2=2.803$ meV}, \mbox{$D=0.635$ meV}, \mbox{$K=0.187$ meV} and  \mbox{$D'=0$ meV}. 
	
	The nonequilibrium electronic spin density which interacts with LMM at site $p$ via the fifth ($sd$ exchange) term on the RHS of Eq.~\eqref{eq:Hcls}
	\begin{equation}\label{eq:noneqspinp}
		\langle \hat{\bf s}_p \rangle (t) = \langle \hat{\bf s}_{i} \rangle (t) + \langle \hat{\bf s}_{j} \rangle(t),
	\end{equation}
	is constructed as the sum of nonequilibrium electronic spin densities on Sn sites $i$ and $j$ with Mn site $p$ being in the middle of the bond between them, as illustrated by the inset in the lower right corner of Fig.~\ref{fig:fig2}. In other words, we assume that both of these electronic spin densities interact with classical LMM of Mn atoms positioned in between sites $i$ and $j$. The nonequilibrium electronic spin densities are obtained as quantum statistical expectation value
	\begin{equation}\label{eq:noneqspinij}
		\langle \hat{\bf s}_i \rangle (t) =  {\rm Tr}_\mathrm{spin}\, [ {\bm \rho}_{ii} \hat{\bm \sigma}],
	\end{equation}
	using  $2 \times 2$ submatrices ${\bm \rho}_{ii}$, composed of elements ${\rho}_{ii,\sigma\sigma'}(t)$, along the diagonal of the matrix representation of the nonequilibrium density operator in the site representation. Its time evolution via QME is explained in Sec.~\ref{sec:lindblad}. 
	
	In the setup of Fig.~\ref{fig:fig1}, we use two MLs of Mn$_3$Sn and one ML of SO material like Pt. Each ML is modeled on a $3 
	\times 3$ lattice (so, there are 54 classical LMMs within Mn$_3$Sn layer composed of two MLs) with periodic boundary conditions employed in both directions using  hoppings between  TB sites in Eq.~\eqref{eq:Htot} on opposite edges, as well as exchange coupling and DMI between classical LMMs in Eq.~\eqref{eq:Hcls} on opposite edges.  
	
	\subsection{Quantum dynamics of electrons with dissipation and decoherence from Lindblad QME}\label{sec:lindblad}
	
	The nonequilibrium density operator is obtained by solving the Lindblad-type~\cite{Lindblad1976,Manzano2020} QME
	\begin{eqnarray}
		\frac{\partial \hat{\rho}}{\partial t} & = & -\frac{i}{\hbar}\big[\hat{H}(t),\hat{\rho}\big] + 	\mathcal{L}[\hat{F}](\hat{\rho}), \label{eq:qme} \\
		\mathcal{L}[\hat{F}](\hat{\rho}) & = & \hat{F}\hat{\rho}\hat{F}^{\dagger} - \frac{1}{2}(\hat{F}^{\dagger}\hat{F}\hat{\rho} + \hat{\rho}\hat{F}^{\dagger}\hat{F}), \label{eq:LSO}
	\end{eqnarray}
	where the first (von Neumann) term on the RHS describes unitary time evolution while the second (Lindblad) term~\cite{Topler2021} on the RHS accounts for dissipation and decoherence effects due to coupling to an external  bosonic bath provided by phonons~\cite{Sold2018}. Since we treat LMMs as the part of the system of interest, magnons and their effect on electrons are included explicitly, rather than being part of external to the system bosonic bath.  The Lindblad superoperator $\mathcal{L}[\hat{F}](\hat{\rho})$ of a jump operator $\hat{F}$ acts on the density operator $\hat{\rho}$ to introduce energy transfer between the electron system and the bath, as well as decoherence of electron system, thereby determining how fast the laser-excited electron system relaxes toward equilibrium. The conservation of the number of electrons is ensured by $\mathrm{Tr}\, \big[ \mathcal{L}[\hat{F}](\hat{\rho}) \big] \equiv 0$.  Without $\mathcal{L}[\hat{F}](\hat{\rho})$ term in Eq.~\eqref{eq:qme}, and for an isolated quantum electronic system within magnetic bilayer in Fig.~\ref{fig:fig1} lacking continuous energy spectrum, pumped charge and spin currents  do not die out even long after the laser pulse has ceased which is obviously nonphysical. Note that 
	continuous energy spectrum is  usually introduced~\cite{Petrovic2018,Bajpai2019a,Bajpai2020,Petrovic2021,Gaury2014,Popescu2016} into quantum transport calculations by attaching finite-size system to semi-infinite leads, but they are not pertinent to light-driven systems [Fig.~\ref{fig:fig1}] used in THz spintronics.
	
	 %----------------------------------------------------------
	\begin{figure}
		\centering
		\includegraphics[width=\linewidth]{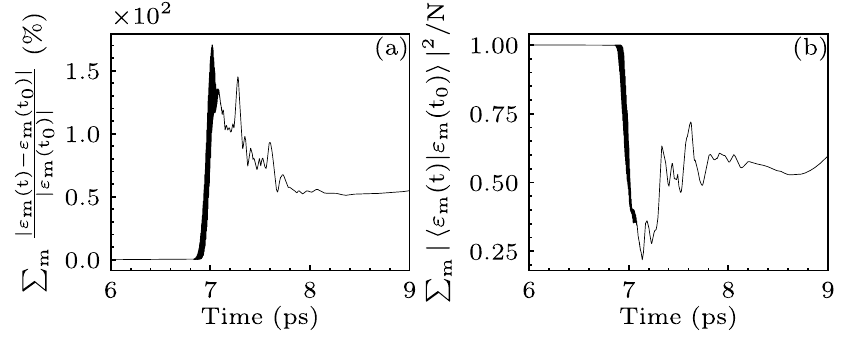}
		\caption{Time dependence of: (a) relative change between instantaneous eigenvalues $\varepsilon_m(t)$ of magnetic bilayer Hamiltonian 
		$\hat{H}_\mathrm{tot}(t)$ [Eq.~\eqref{eq:Htot}] during and after the application of the laser pulse  and eigenvalues $\varepsilon_m(t_0)$ of the same Hamiltonian in equilibrium; and (b) the sum of modulus square of the overlap between the corresponding eigenvectors  $\ket{\varepsilon_m(t)}$ and $\ket{\varepsilon_m(t_0)}$ divided by the total number of eigenstates $N$.}
		\label{fig:fig6}
	\end{figure}
	%----------------------------------------------------------
	
	The jump operator $\hat{F}_{nm} \equiv |\varepsilon_n(t) \rangle \langle \varepsilon_m(t)|$ mediates an inelastic transition (when $\Delta \varepsilon \neq 0$) from eigenstate $\ket{\varepsilon_m(t)}$ with energy $\varepsilon_m(t)$ to eigenstate $\ket{\varepsilon_n(t)}$ with eigenenergy $\varepsilon_n(t)$, thereby transferring energy  \mbox{$\Delta \varepsilon = \varepsilon_n(t) - \varepsilon_m(t)$} from the bath to the electron system for  $\Delta \varepsilon >0$ or vice versa for $\Delta \varepsilon <0$. This operator is 
	expressed using the instantaneous energy eigenbasis $\ket{\varepsilon_m(t)}$ 
	\begin{equation}\label{eq:insteigen}
		\hat{H}_\mathrm{tot}(t) \ket{\varepsilon_m(t)} =\varepsilon_m(t) \ket{\varepsilon_m(t)}, 
	\end{equation}
	instead of static eigenbasis (employed in Ref.~\cite{Topler2021}) of Hamiltonian $\hat{H}_\mathrm{tot}(t<0)$ before application of the laser pulse. Our choice is more realistic as laser pulse, as well as 
	subsequent change in the orientation of LMMs, can {\em dramatically} change energy spectrum of electronic system. This feature can be quantified by computing time-dependent spectral functions~\cite{Gillmeister2020}, or in our case simply by comparing [Fig.~\ref{fig:fig6}(a)] that the relative change \mbox{$\sum_{m=1}^N |\varepsilon_m(t) - \varepsilon_m(t_0)|/|\varepsilon_m(t_0)|$}, between instantaneous eigenvalues $\varepsilon_m(t)$ of magnetic bilayer Hamiltonian 
	$\hat{H}_\mathrm{tot}(t)$ [Eq.~\eqref{eq:Htot}] during the application of the laser pulse and eigenvalues $\varepsilon_m(t_0)$ of the same Hamiltonian in equilibrium,  can reach  $>100\%$ during the application of the laser pulse. Concurrently, the overlap between the corresponding eigenstates $\ket{\varepsilon_m(t)}$ and  $\ket{\varepsilon_m(t_0)}$ drops from $1$ to $\simeq 0.25$ [Fig.~\ref{fig:fig6}(b)]. Note that substantial changes in the  band structure of ultrafast-light-driven magnetic materials have also  been confirmed experimentally~\cite{Gillmeister2020,Eich2017}. 
	
	The importance of our choice is further corroborated by all interlayer currents eventually going  to zero (movie  {\tt equilibration.mp4} in the SM~\cite{sm}) in the equilibration phase (Sec.~\ref{sec:eqdynamics}) at the beginning of which we allow LMMs to interact with room temperature electrons; or after the laser pulse has ceased (movie {\tt after\_laser\_pulse.mp4} in the SM~\cite{sm}). Note that intralayer local spin currents {\em can be nonzero} [Figs.~\ref{fig:fig4}(d)--(f)] even in equilibrium, that is before the light pulse is applied, due to  SOC~\cite{Nikolic2006,Rashba2003,Droghetti2022} present in either Mn$_3$Sn or Pt layer. Indeed, if 
	all SOCs are turned off [gray curves in Figs.~\ref{fig:fig4}(d) and ~\ref{fig:fig4}(e)], spin currents are zero  before the light pulse is applied. Conversely, we find that the choice of static eigenbasis from  Ref.~\cite{Topler2021} to define Lindbladian leads to nonphysical situation with nonvanishing charge currents in equilibrium or after the laser pulse has ceased.  
	
	The Lindbladian $\mathcal{L}[\hat{F}_{nm}](\hat{\rho})$ is then given by
	\begin{equation}\label{eq:L1}
		\begin{split}
			\hat{L}_{nm}(\hat{\rho}) \equiv &\mathcal{L}[\hat{F}_{nm}](\hat{\rho})= \ket{\varepsilon_n(t)}\rho_{mm}(t)\bra{\varepsilon_n(t)}\\
			&-\frac{1}{2}\sum_{k}\big( \ket{\varepsilon_m(t)}\rho_{mk}(t)\bra{\varepsilon_k(t)} \\
			&+ \ket{\varepsilon_k(t)}\rho_{km}(t)\bra{\varepsilon_m(t)} \big),
		\end{split}
	\end{equation}
	where $\rho_{nm}(t) = \bra{\varepsilon_n(t)}\hat{\rho} (t)\ket{\varepsilon_m(t)}$ are the matrix elements of nonequilibrium density operator in the basis of instantaneous energy eigenstates. Finally, the total Lindbladian
	\begin{equation}\label{eq:L2}
		\hat{L}\big(\hat{\rho}(t)\big) = \sum_{nm} \Gamma_{nm}(t)\hat{L}_{nm}\big( \hat{\rho}(t) \big),
	\end{equation}
	is  a weighted sum over all $\hat{L}_{nm}$. The functions $\Gamma_{nm}(t)$ are given by~\cite{Topler2021}
	\begin{equation}\label{eq:gamma}
		\Gamma_{nm}(t) = 
		\begin{cases}
			\Gamma_\Box \pi_{nm}(t)[f_\mathrm{BE}(\Delta \varepsilon, \mu, T)+1]\qquad &\mathrm{for} \ \mathrm{\Delta \varepsilon<0},\\
			\Gamma_\Box \pi_{nm}(t)[f_\mathrm{BE}(\Delta \varepsilon, \mu, T)] &\mathrm{for} \ \mathrm{\Delta\varepsilon>0},\\
			\Gamma_{\rm PD} &\mathrm{for} \ \mathrm{\Delta\varepsilon=0},
		\end{cases}
	\end{equation}
	with $\Box$ being a placeholder for either spin-conserving (SC) or spin-flip (SF) subscript. Here \mbox{$\pi_{nm}(t) = \rho_{mm}(t)[1-\rho_{nn}]$} factors take care of the Pauli exclusion principle for fermionic statistics~\cite{HeadMarsden2015}, as well as ensuring \mbox{$0 \leq \rho_{nn}(t) \leq 1$} for each eigenstate and at all times; and $f_\mathrm{BE}$ is the Bose-Einstein distribution function (with chemical potential $\mu=0$) of the bath at temperature $T$ (set to \mbox{$T = 300$ K}). Thus, for \mbox{$n \neq m$}, Lindbladian $\hat{L}_{nm}(\hat{\rho})$ generates effective thermalization of electronic subsystem on the typical timescale of \mbox{$\sim 1$ ps}, e.g., we use \mbox{$\Gamma_{\rm SC}=2\times10^{-4}$ fs$^{-1}$} corresponding to timescale of \mbox{$5$ ps} (as also employed in Ref.~\cite{Topler2021}). Conversely, for \mbox{$n = m$}, $\hat{L}_{nn}(\hat{\rho})$ account for ``true decoherence''~\cite{Joos2003} or ``pure dephasing'' (PD)~\cite{Chen2019b}  which reduces the off-diagonal elements of $\hat{\rho}(t)$ with the rate $\Gamma_{\rm PD}$. ``Pure dephasing'' can occur even when no energy is exchanged between the system and the environment, so $\mathrm{\Delta\varepsilon=0}$ in Eq.~\eqref{eq:gamma},  as exemplified by microscopic processes where the system is entangled to an environment having degenerate energy eigenstates~\cite{Joos2003}. We use \mbox{$\Gamma_{\rm PD}=5\times10^{-2}$ fs$^{-1}$}, which correspond to timescales of \mbox{$20$ fs}. 	
	
	 %----------------------------------------------------------
	\begin{figure}
		\centering
		\includegraphics[width=\linewidth]{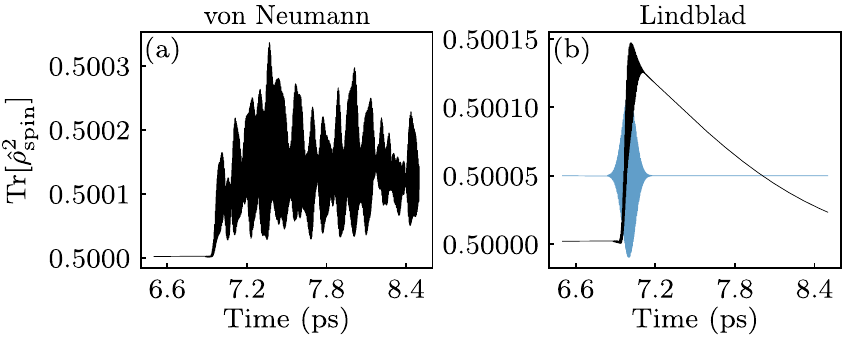}
		\caption{Time dependence of purity~\cite{Ziolkowski2023}, $\mathrm{Tr}\, \hat{\rho}_\mathrm{spin}^2$, of spin density matrix [Eq.~\eqref{eq:spindensitymatrix}] of electronic subsystem within magnetic bilayer in Fig.~\ref{fig:fig1} under: (a) unitary (von Neumann) time evolution via Eq.~\eqref{eq:qme} with $\mathcal{L}[\hat{F}](\hat{\rho})=0$; and (b) Lindblad time evolution via Eq.~\eqref{eq:qme}. For reference, the blue curve in panel (b) depicts the time frame within which fs laser pulse (with arbitrary units on its ordinate) is applied.}
		\label{fig:purity}
	\end{figure}
	%----------------------------------------------------------
	
	We also set \mbox{$\Gamma_{\rm SF}=0$} because in the presence of SO coupling electron eigenstates do not have well-defined spin due to {\em entanglement} of spin and orbital factor states~\cite{Nikolic2005,Stav2018,Gotfryd2020}. That is, eigenstates of SO-coupled electronic subsystem can be written (in Schmidt decomposition~\cite{Ekert1994}) as 	$\ket{E_n} = \ket{\Phi_n^\prime} \otimes \ket{\uparrow} + \ket{\Phi_n^{\prime \prime}} \otimes \ket{\downarrow}$, which prevents defining SF transitions (with the rate \mbox{$\Gamma_{\rm SF} \neq 0$}) between eigenstates $\ket{E_n^\prime} = \ket{\Phi_n^\prime} \otimes \ket{\uparrow}$ and $\ket{E_n^{\prime\prime}} = \ket{\Phi_n^{\prime \prime}} \otimes \ket{\downarrow}$ in the absence if SO coupling which have well-defined spin. Nevertheless, this does not mean that spin dephasing is not included in our calculations. For example, Fig.~\ref{fig:purity} shows purity~\cite{Ziolkowski2023} $\mathrm{Tr}\, \hat{\rho}_\mathrm{spin}^2$ of electronic spin density matrix, 
	\begin{equation}\label{eq:spindensitymatrix}
	\hat{\rho}_\mathrm{spin} = \mathrm{Tr}_\mathrm{orbital} \hat{\rho},
	\end{equation}
	obtained by partial tracing of the full density matrix $\hat{\rho}$ (interestingly, purity $\mathrm{Tr}\, \hat{\rho}^2$  of $\hat{\rho}$ for ultrafast-light-driven magnetic bilayers was studied very recently in Ref.~\cite{Ziolkowski2023}) over the orbital states. Purity equal to one signifies electronic spin being in pure quantum state, $\ket{\Sigma}$ or \mbox{$\hat{\rho}_\mathrm{spin}=\ket{\Sigma}\bra{\Sigma}$}, while its value below 1 in equilibrium or in the course of unitary (von Neumann) evolution [Fig.~\ref{fig:purity}(a)], is due to mixed nature of spin states in SO-coupled systems. Its further decay [Fig.~\ref{fig:purity}(b)] upon switching to Lindbladian dynamics via Eq.~\eqref{eq:qme} confirms that spin dephasing is intrinsically included in our scheme defining [Eq.~\eqref{eq:gamma}] action of the Lindblad operators between energy eigenstates of SO-coupled Hamiltonian. 
	
	The outcome of solving Eq.~\eqref{eq:qme} at each time step $\delta t =0.1$ fs is the density matrix ${\rho}_{mn}(t)$ represented in the instantaneous energy eigenbasis [Eq.~\eqref{eq:insteigen}]. Once transformed back to site basis, its ${\bm \rho}_{ii}(t)$ $2 \times 2$ submatrices determine [Eq.~\eqref{eq:noneqspinij}] the nonequilibrium spin density $\langle \hat{\bf s}_p \rangle$  entering as spin torque $\mathbf{T}_p = J_{sd} \langle \hat{\bf s}_p\rangle(t) \times \mathbf{M}_p(t)$ into the LLG equation (Sec.~\ref{sec:llg}) part of the loop in Fig.~\ref{fig:fig2} via the fifth term on the RHS of classical Hamiltonian [Eq.~\eqref{eq:Hcls}] of LMMs. In turn, the LLG equation updates the orientation of LMMs $\mathbf{M}_p(t)$ and, therefore, modifies time-dependence of the third term on the RHS of Eq.~\eqref{eq:H1}. Finally, at the same instant of time in which ${\bm \rho}_{ij}(t)$ and $\mathbf{M}_{ij}(t)$ are self-consistently computed, we also compute bond~\cite{Nikolic2006,Petrovic2018,Ridley2021} charge
	\begin{equation}\label{eq:bondcharge}
		I_{i\rightarrow j}(t) = \frac{e\gamma}{i\hbar}\mathrm{Tr}_{\rm spin}\bigg[{\bm \rho}_{ij}(t)\mathbf{H}_{ji}(t) - {\bm \rho}_{ji}(t)\mathbf{H}_{ij}(t)\bigg],
	\end{equation}
	and spin currents
	\begin{equation}\label{eq:bondspin}
		I_{i\rightarrow j}^{S_\alpha}(t) = \frac{e\gamma}{i\hbar}\mathrm{Tr}_{\rm spin}\bigg[\hat{\sigma}_\alpha\bigg\{{\bm \rho}_{ij}(t)\mathbf{H}_{ji}(t) - {\bm \rho}_{ji}(t)\mathbf{H}_{ij}(t)\bigg\}\bigg],
	\end{equation}
	from the off-diagonal $2 \times 2$ submatrices  ${\bm \rho}_{ij}(t)$. The bond charge current  $I_{i\rightarrow j}(t)$ is plugged into the Jefimenko equations (Sec.~\ref{sec:jefimenko}), together with time-dependent on-site charge density 
	\begin{equation}\label{eq:onsitecharge}
		e \rho_{ii}(t) = \mathrm{Tr}_{\rm spin}[{\bm \rho}_{ii}(t)], 
	\end{equation}
	in order to obtain time-dependent electric $\mathbf{E}(t)$ and magnetic $\mathbf{B}(t)$ fields of outgoing EM radiation at time $t$. Equations~\eqref{eq:bondcharge} and ~\eqref{eq:bondspin} show that the larger the off-diagonal elements of the 
	nonequilibrium density matrix in site representation (due to laser pulse or LMM dynamics) and hopping between the sites [modified by the 
	laser pulse via Eq.~\eqref{eq:peierls}], the larger currents will be generated. 
	
\begin{figure*}
		\centering
		\includegraphics[width=\linewidth]{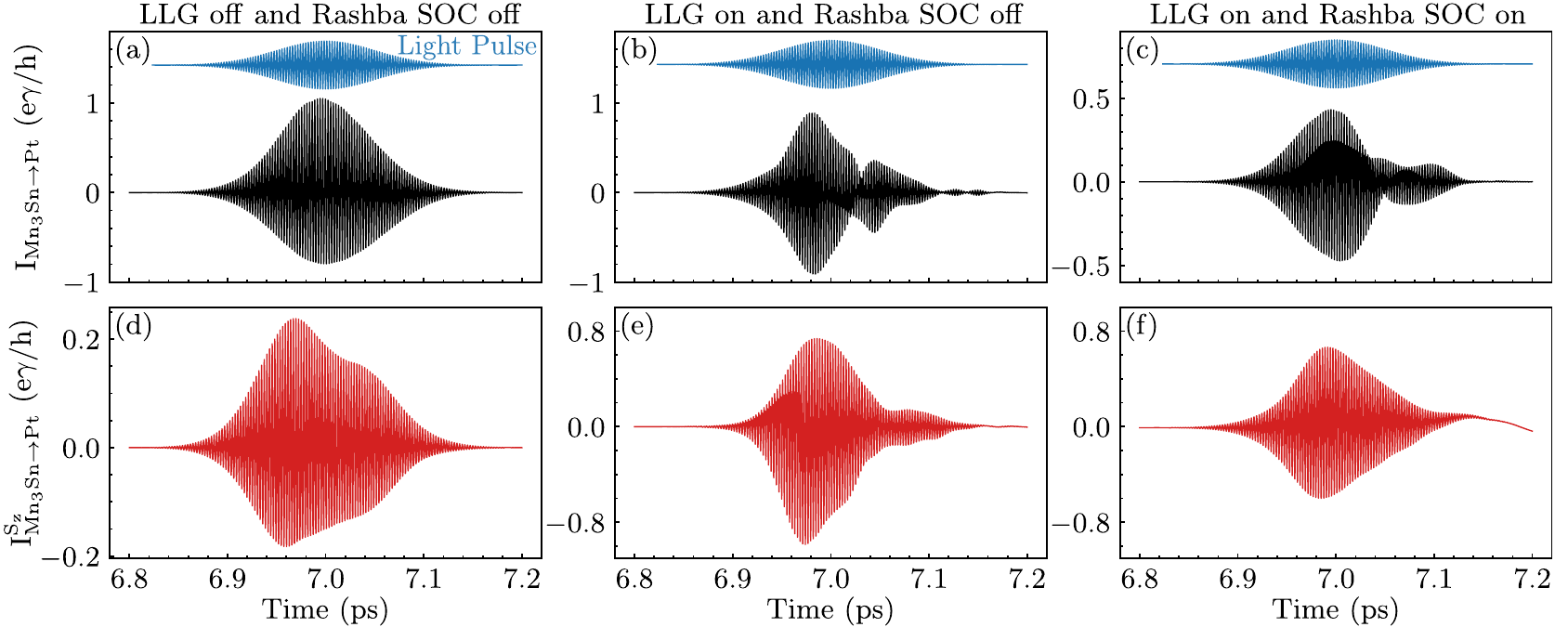}
		\caption{Time dependence of the {\em sum} of (a)--(c) all interlayer charge, $I_{\mathrm{Mn}_3\mathrm{Sn} \rightarrow \mathrm{Pt}}$, and (d)--(f) spin-$z$,  $I^{S_z}_{\mathrm{Mn}_3\mathrm{Sn} \rightarrow \mathrm{Pt}}$, local (or  bond, i.e., between two sites of TB lattice~\cite{Nikolic2006,Petrovic2018,Ridley2021}) currents, initiated by fs laser  pulse irradiation of Mn$_3$Sn layer in Fig.~\ref{fig:fig1}. These interlayer currents flow along the $z$-axis in the coordinate system of Fig.~\ref{fig:fig1}. We artificially turn off: (a) and (d) both LLG dynamics of LMMs within Mn$_3$Sn (so LMMs are static and time-independent) layer and SOC with Pt layer; and (b) and (e) SOC within Pt layer. In panels (c) and (f) all terms in quantum [Eq.~\eqref{eq:Htot}] and classical [Eq.~\eqref{eq:Hcls}] Hamiltonians are active.  For reference, the blue curve on top of panels (a)--(c) depicts the time frame within which fs laser pulse (with arbitrary units on its ordinate) is applied.}
		\label{fig:fig3}
	\end{figure*}

	Note that total interlayer or intralayer charge and spin  currents plotted in Figs.~\ref{fig:fig3}--\ref{fig:fig4} are obtained by summing all respective charge and spin bond current contributions. For example,  Fig.~\ref{fig:fig4}(a)--(c) plots total intralayer charge current $I_{\mathrm{Mn}_3Sn} + I_{\mathrm{Pt}} = \sum_{\langle i,j \rangle} I_{i\rightarrow j}$ with \mbox{${\langle i,j \rangle} \in$ Mn$_3$Sn} or  \mbox{${\langle i,j \rangle} \in$ Pt} and  $\mathbf{e}_{ij} \parallel \mathbf{e}_x$, where   $\mathbf{e}_x$ is the unit vector along the $x$-axis and  $\mathbf{e}_{ij}$ is the unit vector connecting sites $i$ and $j$.
	
	%{Bond current is calculated in \mbox{particle~fs$^{-1}$}. After multiplying with 2$\pi$ the unit of current is $e\gamma/h$ equivalent to electronic charge-fs$^{-1}$. So $\rm val/0.65829$ is in $\rm[efs^{-1}]$, which in SI units is $\rm (val/0.65829)\times1.60217\times10^{-4}\ A$. Now we use SI units to proceed with calculation of emitted radiation.%}

	\subsection{Classical dynamics of LMMs from LLG equation}\label{sec:llg}
	
	The classical LMMs are evolved by solving the LLG equation
	\begin{equation}\label{eq:llg}
		\frac{\partial\mathbf{M}_{p}}{\partial t} =
		-\frac{g}{1 + {\lambda}^{2}}\left[\mathbf{M}_{p} \times \mathbf{B}^{\rm eff}_{p} 
		+\lambda \mathbf{M}_{p} \times \left(\mathbf{M}_{p} \times \mathbf{B}^{\rm eff}_{p}\right)\right],
	\end{equation}
	via the Heun numerical scheme with projection onto the unit sphere~\cite{Evans2014} and time step \mbox{$\delta t =0.1$ fs}. Here \mbox{$\mathbf{B}_{\rm eff}^{p} = - \frac{1}{\mu_M} \partial \mathcal{H} /\partial \mathbf{M}_{p}$}  is the effective magnetic field ($\mu_M$ is the magnitude of LMMs); $g$ is the gyromagnetic ratio; and $\lambda$ is Gilbert damping. Note that in the case of Weyl conduction electrons surrounding classical LMMs,  \mbox{$\lambda=0$}~\cite{Nikolic2021} when all LMMs precess uniformly as a macrospin. Since LMMs within Mn$_3$Sn are noncollinear $\lambda \neq 0$, but wide range of values has been extracted from experiments (such as $\lambda \sim 10^{-4}$~\cite{Miwa2021}) or assumed in numerical simulations  (such as $\lambda \sim 10^{-2}$~\cite{Nomoto2020}). Here we use \mbox{$\lambda = 0.2$} for convenience, so that LMMs are quickly brought into equilibrium on a reasonable time scale after their dynamics is generated by coupling them to Weyl electrons at $t=0$, as animated in the movie {\tt equilibration.mp4}  in the SM~\cite{sm}. This ensures that LMMs do not move and no currents are pumped by their motion prior to the introduction of laser pulse.
	
	%----------------------------------------------------------
	%----------------------------------------------------------
	
	%==========================================================
	\subsection{Retarded electric and magnetic field of radiation from Jefimenko equations}\label{sec:jefimenko}
	
	The emitted EM radiation is calculated from the Jefimenko equations~\cite{Jefimenko1966} for the retarded electric 
	\begin{equation}\label{eq:efield}
		\begin{split}
			&{\bf E}(\mathbf{r},t) = \frac{1}{4\pi\epsilon_0}\bigg\{\sum_{i=1}^{N}  \bigg[e\rho_{ii}(t_r) \frac{\mathbf{r}- \mathbf{r}_i}{|\mathbf{r}-\mathbf{r}_i|^3} + \frac{1}{c}\frac{\partial e\rho_{ii}(t_r)}{\partial t}\frac{\mathbf{r} - \mathbf{r}_i}{|{\bf r}-{\bf r}_i|^2}\bigg]\\
			&\ \ \ \ \ - \sum_{P_{i \rightarrow j}=1}^{N_b} \int_{P_{i \rightarrow j}} \frac{1}{c^2}\frac{\partial}{\partial t}\frac{I_{i \rightarrow j}(t_r)}{|{\bf r}-{\bf l}|}d{\bf l} \bigg\},
		\end{split}
	\end{equation}
	and retarded magnetic fields
	\begin{equation}\label{eq:bfield}
		\begin{split}
			{\bf B}(\mathbf{r},t) =&\ \frac{\mu_0}{4\pi}\sum_{P_{i \rightarrow j}=1}^{N_b} \int_{P_{i \rightarrow j}} \bigg[\frac{I_{i \rightarrow j}(t_r)}{|{\bf r}-{\bf l}|^3}\\&+ \frac{1}{c}\frac{\partial}{\partial t}\frac{I_{i \rightarrow j}(t_r)}{|{\bf r}-{\bf l}|^2}\bigg]d{\bf l}\times ({\bf r}-{\bf l}),
		\end{split}
	\end{equation}
	where these equations are adapted~\cite{Ridley2021} to use time-dependent bond charge currents $I_{i \rightarrow j}(t)$ [Eq.~\eqref{eq:bondcharge}] and on-site electronic charge density $e\rho_{ii}(t)$ [Eq.~\eqref{eq:onsitecharge}] defined on the TB lattice. The bond currents~\cite{Nikolic2006,Petrovic2018,Ridley2021} $I_{i \rightarrow j}$ are assumed to be spatially homogeneous along the path $P_{i \rightarrow j}$ from site $i$ to site $j$, which is composed of a set of points $\mathbf{l} \in P_{i \rightarrow j}$. The on-site charge density $e\rho_{ii}(t)$ and bond charge currents $I_{i \rightarrow j}(t)$ satisfy the continuity equation~\cite{Ridley2021}
	\begin{equation}\label{eq:continuity}
		e	\frac{\partial \rho_{jj}(t)}{\partial t} = \sum_i I_{i \rightarrow j}(t).
	\end{equation}
	In Eqs.~\eqref{eq:efield} and \eqref{eq:bfield} we have $N=27$ as the total number of TB sites; $N_b=108$ as the total number of bonds between them; and \mbox{$t_r \equiv t -|\mathbf{r}-\mathbf{l}|/c$} emphasizes retardation in the response time due to relativistic causality. Equations~\eqref{eq:efield} and ~\eqref{eq:bfield} are applicable in approximation where we neglect self-consistent effects, such as emitted EM radiation exerting backaction onto magnetic bilayer in Fig.~\ref{fig:fig1} (capturing such effects would require~\cite{Philip2018} solving Maxwell equations for the scalar and vector electromagnetic potential, which can then be plugged into the Hamiltonian in Eq.~\eqref{eq:Htot}, thereby establishing extra self-consistent loop in Fig.~\ref{fig:fig2}). 

	We note that details of charge current and charge density spatial profiles within nanoscale thickness of THz emitters, such as bilayer in Fig.~\ref{fig:fig1}, are not important~\cite{Nenno2019} for computing emitted THz radiation with wavelengths in the micrometer to submillimeter range. Nevertheless, since we compute in Fig.~\ref{fig:fig5} emitted EM radiation in both THz range and for nm range wavelengths of  incoming light and its high harmonics, we keep full spatial profile of $e \rho_{ii}(t)$ and $I_{i \rightarrow j}(t)$ in the Jefimenko Eqs.~\eqref{eq:efield} and ~\eqref{eq:bfield} without any simplifications.

	\section{Results and Discussion}\label{sec:results}

	\subsection{Equilibrium dynamics initiated by coupling LMMs to room temperature Weyl electrons}\label{sec:eqdynamics}
	
	The hexagonal-type lattice (space group $P6_3/mmc$) of Mn$_3$Sn is comprised of kagome lattice of Mn magnetic atoms with geometrical frustration leading to a noncollinear (i.e., inverse triangular) configuration of LMMs with antiferromagnetic ordering~\cite{Yang2017,Liu2017}. Thus, unlike standard collinear antiferromagnetic order of LMMs, the noncollinear configuration of LMMs of Mn atoms breaks time-reversal symmetry (TRS) macroscopically. 
	
	For times $t<0$, we put LMMs of Mn in the  noncollinear configuration depicted in Fig.~\ref{fig:fig1}, and concurrently we describe electrons by the grand canonical equilibrium density operator   
	\begin{equation}\label{eq:initialdm}
		\hat{\rho}_{\rm eq} = f_\mathrm{FD}(\hat{H}_\mathrm{tot} -E_F \hat{I}),
	\end{equation}
	at room temperature \mbox{$T=300$ K}, where \mbox{$f_\mathrm{FD}(x)=(1+e^{\beta x})^{-1}$} is the Fermi-Dirac distribution function; $\beta = 1/k_B T$;  the Fermi energy $E_F$ is chosen at half filling of TB lattice of bilayer in Fig.~\ref{fig:fig1}; and $\hat{I}$ is the unit operator in the Hilbert space of electrons.  At $t=0$, we  couple room temperature conduction electrons to classical LMMs by switching on $J_{sd}(t \ge 0) \neq 0$ in Eqs.~\eqref{eq:Htot} and ~\eqref{eq:Hcls}, and then evolve both subsystems using QME+LLG self-consistent loop within Fig.~\ref{fig:fig2}. Due to the Lindblad terms in electronic 
	equation of motion [Eq.~\eqref{eq:qme}] and the Gilbert damping term [Eq.~\eqref{eq:llg}] in the LLG equation of motion for classical LMMs, this evolution over \mbox{$2$ ps} time interval reaches equilibrium signified by all transient charge and spin currents decaying to {\em zero}, as animated by the movie {\tt equilibration.mp4} in the SM~\cite{sm}. The resulting noncollinear configuration of LMMs of Mn atoms, which will interact with light-driven electrons once the laser pulse is applied, is randomized (movie {\tt equilibration.mp4} in the SM~\cite{sm}) with respect to original inverse triangular configuration depicted in Fig.~\ref{fig:fig1}. It also possesses small nonzero local magnetization per site (panel $M^\alpha/N_\mathrm{LMM}$ in the movie  {\tt equilibration.mp4} in the SM~\cite{sm}).

	%----------------------------------------------------------
	\begin{figure*}
		\centering
		\includegraphics[width=\linewidth]{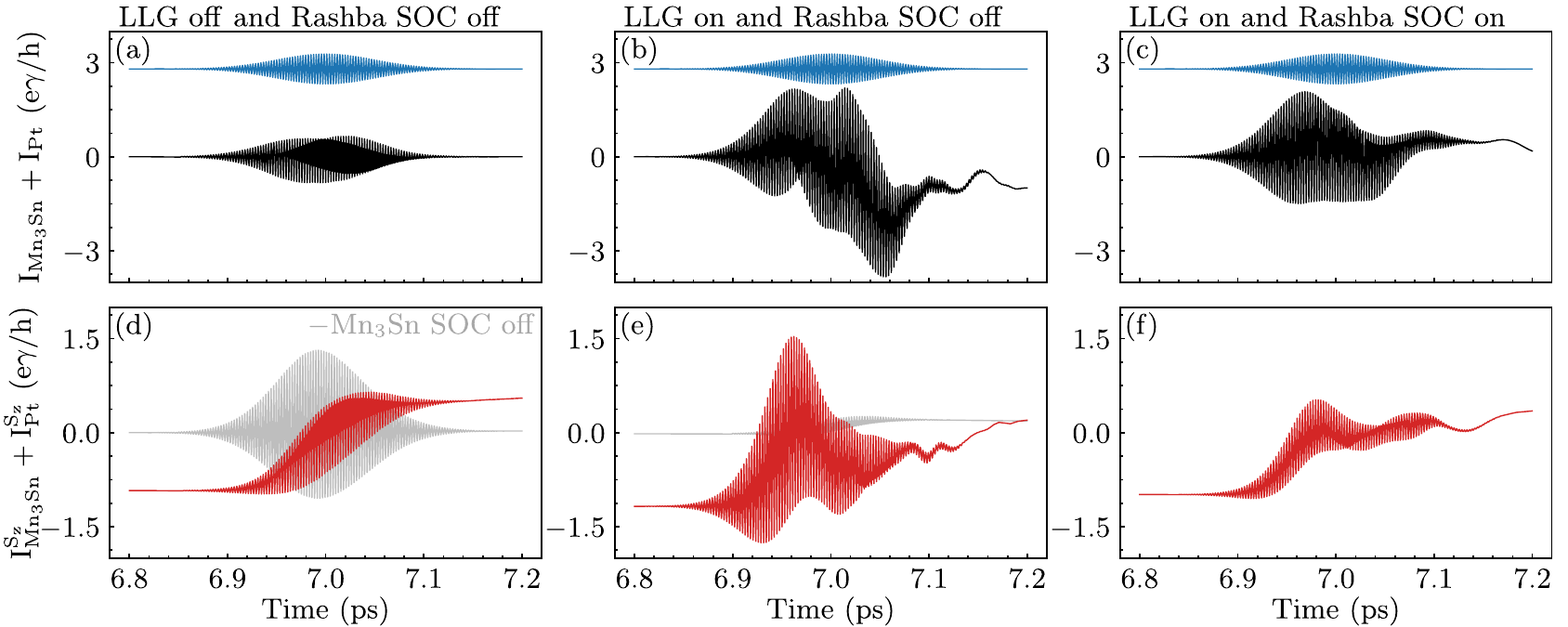}
		\caption{The same information as in Fig.~\ref{fig:fig3}, but for the {\em sum} of (a)--(c) all intralayer charge, $I_{\mathrm{Mn}_3\mathrm{Sn}} + I_\mathrm{Pt}$, and (d)--(f) spin-$z$,  $I_{\mathrm{Mn}_3\mathrm{Sn}}^{S_z} +  I_\mathrm{Pt}^{S_z}$, local (or  bond, i.e., between two sites of TB lattice~\cite{Nikolic2006,Petrovic2018,Ridley2021}) currents within the whole bilayer in Fig.~\ref{fig:fig1}. These intralayer currents flow along the $x$-axis in the coordinate system of Fig.~\ref{fig:fig1}.  Light gray curves in panels (d) and (e) are obtained by additionally switching off SOC within Mn$_3$Sn layer. For reference, the blue curve on top of panels (a)--(c) depicts the time frame within which fs laser pulse (with arbitrary units on its ordinate) is applied.}
		\label{fig:fig4}
	\end{figure*}
	%----------------------------------------------------------

	\subsection{Spin and charge currents pumped by light and/or local magnetization dynamics}\label{sec:currents}

	After equilibration procedure from Sec.~\ref{sec:eqdynamics} is completed, the electronic subsystem is excited by shining fs laser pulse onto both MLs of Mn$_3$Sn layer. The laser pulse is described by the vector potential 
	\begin{equation}\label{eq:vectorpot}
		\mathbf{A}(t) = A_\mathrm{max} e^{-\frac{(t-t_p)^2}{2\sigma^2}} \cos(\Omega_0 t) \mathbf{e}_x,
	\end{equation}
	with a Gaussian shaped function for a pulse of duration $\sigma$, centered at time $t_p$ and with center frequency $\Omega_0$. Here $\mathbf{e}_x$ accounts  for one the two possible linear polarizations within the $xy$-plane of incident light. The corresponding electric field is \mbox{$\mathbf{E}=-\partial \mathbf{A}/\partial t$}. The relativistic magnetic field of the laser pulse affects electronic spin degree, 
	but this effect is negligible and usually not considered~\cite{Wang2017a,Bajpai2019}. The vector potential couples to an electron via the Peierls substitution~\cite{Panati2003,Li2020} in Eq.~\eqref{eq:H1} 
	\begin{subequations}\label{eq:peierls}
	\begin{eqnarray}
		\gamma^{xy}_{ij}(t) & =  & \gamma \exp \bigg \{ \textit{i}z_\mathrm{max} e^{\frac{-(t-t_p)^2}{2\sigma^2}}\cos(\Omega_0 t) \mathbf{e}_x \cdot \mathbf{e}_{ij}  \bigg \} \\
		\gamma^z_{ij}(t) & = & \gamma^z \exp \bigg \{ \textit{i}z_\mathrm{max} e^{\frac{-(t-t_p)^2}{2\sigma^2}}\cos(\Omega_0 t) \mathbf{e}_x \cdot \mathbf{e}_{ij} \bigg \}
	\end{eqnarray}
    \end{subequations}
	where \mbox{$z_\mathrm{max} = e a_0 A_\mathrm{max}/\hbar = 0.12$} is the dimensionless parameter~\cite{Bajpai2019} quantifying maximum amplitude of the pulse; \mbox{$\sigma = 25$ fs} is the width of the Gaussian envelope; and center frequency \mbox{$\Omega_0=2.354$ fs$^{-1}$} is chosen to correspond to \mbox{$800$ nm} wavelength commonly employed~\cite{Rouzegar2022,Seifert2022,Bull2021,Wu2021,Seifert2016,Wu2016,Chen2018,Zhou2019b,Qiu2020,Rongione2022} in THz spintronics. 
	
	%----------------------------------------------------------
	\begin{figure*}
		\centering
		\includegraphics[width=\linewidth]{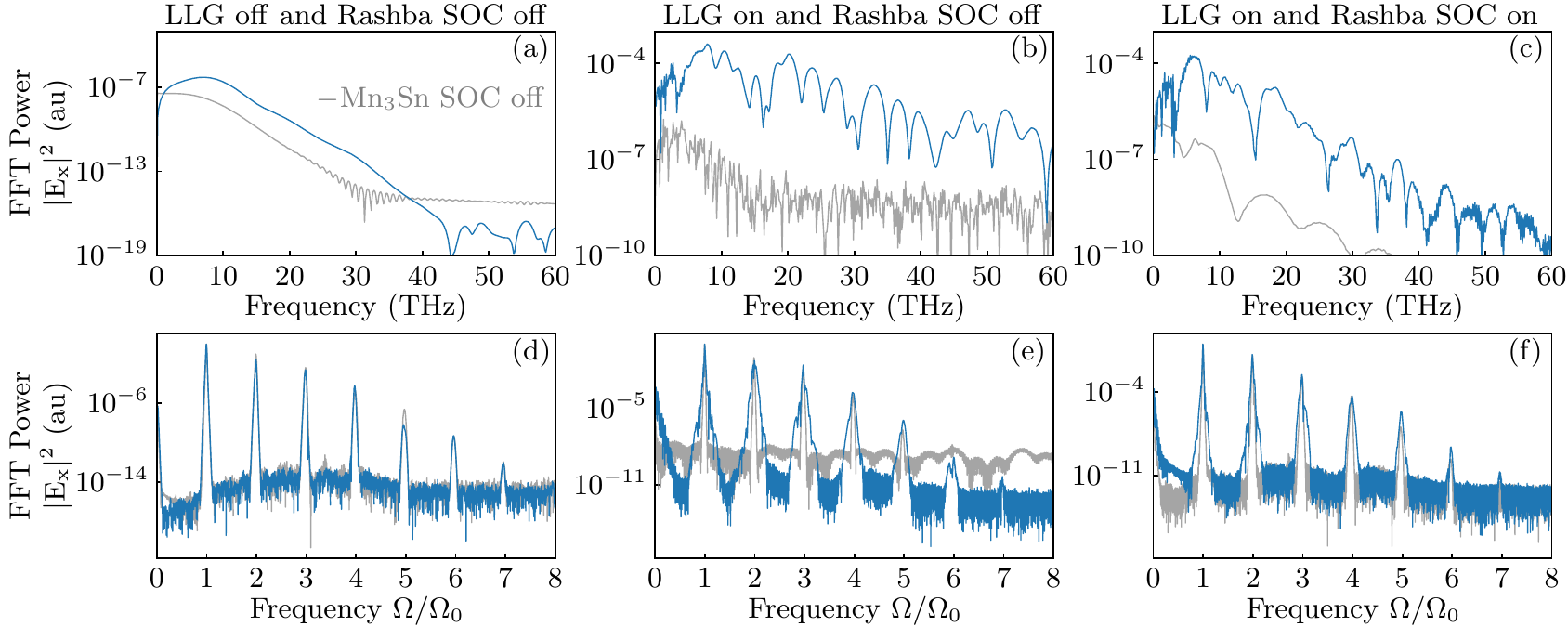}
		\caption{FFT power spectrum of the $x$-component of the electric field $|E_{x}(\mathbf{r},\Omega)|^2$ of emitted EM radiation  calculated at point \mbox{$\mathbf{r} =(0,0,100)a_0$} along the $z$-axis in Fig.~\ref{fig:fig1}. This outgoing EM radiation, computed from the Jefimenko~\cite{Jefimenko1966,Ridley2021} equations [Eqs.~\eqref{eq:efield} and ~\eqref{eq:bfield}] using charge and current  densities obtained from QME+LLG loop in Fig.~\ref{fig:fig2}, is response to incoming fs laser pulse irradiating Mn$_3$Sn layer [Fig.~\ref{fig:fig1}].  Panels (a)---(c) show the FFT power spectrum of EM radiation in the THz range of frequencies, while panels (d)---(f) show the FFT power spectrum of emitted radiation exhibiting high harmonics~\cite{Ghimire2019} of the center frequency $\Omega_0$ of the pulse. Note that the results in the frequency range of panels (a)--(c) are hidden on the left edge of panels (d)--(f). We artificially switch off: (a) and (d) both LLG dynamics of LMMs within Mn$_3$Sn (so LMMs remain time-independent) layer and the Rashba SOC with Pt layer; and (b) and (e) SOC within Pt layer. In panels (c) and (f) all terms in quantum [Eq.~\eqref{eq:Htot}] and classical [Eq.~\eqref{eq:Hcls}] Hamiltonians are active. Light gray curves in (a)--(f) are obtained by additionally switching off SOC within Mn$_3$Sn layer.}
		\label{fig:fig5}
	\end{figure*}
	%----------------------------------------------------------
	
	In the first column of panels in Figs.~\ref{fig:fig3} and \ref{fig:fig4}, we switch off the LLG dynamics and set the Rashba SOC in the NM layer to zero. This means that LMMs are fixed in their configuration reached upon equilibration with room temperature electrons as discussed in Sec.~\ref{sec:eqdynamics}. Thus, hot electrons are flowing due to laser pulse {\em only} to comprise pumped interlayer [Fig.~\ref{fig:fig3}] and intralayer [Fig.~\ref{fig:fig4}] charge [top row of panels in Figs.~\ref{fig:fig3} and ~\ref{fig:fig4}] and spin [bottom row of panels in Figs.~\ref{fig:fig3} and ~\ref{fig:fig4}] currents.  
		
	In the second column of panels in Figs.~\ref{fig:fig3} and \ref{fig:fig4}, we switch on the LLG dynamics while keeping the Rashba SOC zero within the NM layer. This change results in several times enhanced amplitude of interlayer [Fig.~\ref{fig:fig3}(e)] and intralayer [Fig.~\ref{fig:fig4}(e)]  spin currents, as well as nearly the same interlayer [Fig.~\ref{fig:fig3}(b)] charge currents. Importantly, switching on LLG dynamics leads to a much larger intralayer [Fig.~\ref{fig:fig4}(b)] charge current which then enhances [Eq.~\eqref{eq:efield}] the $E_x$-component of electric field of emitted radiation [Fig.~\ref{fig:fig5}(b)].  Thus, comparison of the first and the second column of panels in Figs.~\ref{fig:fig3} and \ref{fig:fig4} reveals how slow dynamics of LMMs, driven by spin current pumped initially only by ultrafast light, provide additional contributions to {\em both} spin and charge currents generation.

	We note that for perfectly collinear LMMs, pumped spin current will have $\langle \hat{\bf s}_{p} \rangle (t) \parallel \mathbf{M}_p$, so that no spin torque  $\mathbf{T}_p = J_{sd} \langle \hat{\bf s}_p \rangle (t) \times \mathbf{M}_p(t)$ is generated in the LLG equation and, therefore, no torque-driven classical dynamics of LMMs is initiated. While this situation is not relevant for the chosen example of light-irradiated Mn$_3$Sn due to noncollinearity of its LMMs in equilibrium, in the case of FM or AF layer with collinear LMMs one should include thermal fluctuations in the initial state of LMMs by, e.g., adding~\cite{Ghosh2022} small random polar and azimuthal angle to the direction of each LMM.
	
	In the third column of panels in Figs.~\ref{fig:fig3} and ~\ref{fig:fig4} and in the movie {\tt during\_laser\_pulse.mp4}, we switch on both the LLG dynamics in Mn$_3$Sn layer and the Rashba SOC in the NM layer. This actually reduces {\em all} pumped spin and charge currents, instead of naively enhancing charge current in NM layer via interfacial spin-to-charge conversion~\cite{Jungfleisch2018a}. Whether interfacial SOC reduces or enhances spin pumping depends on specific materials combination, as found in first-principles quantum transport studies~\cite{Dolui2020a,Dolui2022} of magnetic bilayers.

	\subsection{Emitted EM radiation: THz and high-harmonic generation}\label{sec:emittedthz}
	
	Figure~\ref{fig:fig5} plots fast Fourier transform (FFT) power spectrum $|E_x(\mathbf{r},\Omega)|^2$ of the $x$-component (i.e., parallel to the interface in Fig.~\ref{fig:fig1}) of electric field [Eq.~\eqref{eq:efield}] of emitted EM radiation at point \mbox{$\mathbf{r}=(0,0,100)a_0$} away from magnetic bilayer. Following the same strategy as in Figs.~\ref{fig:fig3} and \ref{fig:fig4}, Fig.~\ref{fig:fig5} is also formatted using three columns of panels where the difference between the first and the second column shows how charge pumping by LMM dynamics enhances radiation in the THz range [Fig.~\ref{fig:fig5}(b) vs. Fig.~\ref{fig:fig5}(a)]. This feature is in full accord with the corresponding Fig.~\ref{fig:fig4}(b), as it can be concluded from the Jefimenko Eq.~\eqref{eq:efield} where $E_x$ component the electric field of emitted EM radiation is generated by intralayer charge currents---of either Mn$_3$Sn or Pt layer with the former providing larger contribution---flowing parallel to the $x$-axis (in the coordinate system of Fig~\ref{fig:fig1}). On the other hand, switching on the Rashba SOC in NM layer reduces the amplitude of THz radiation [Fig.~\ref{fig:fig5}(c)], as expected from reduced charge pumping in Figs.~\ref{fig:fig3}(c) and ~\ref{fig:fig4}(c). 
	
	To clarify the principal contribution to charge current responsible for EM radiation in the THz range, we switch off intrinsic SOC  of Mn$_3$Sn [i.e., $\lambda_z=0$ in the fourth term on the RHS of Eq.~\eqref{eq:H1}]. This trick reveals that SOC is not important for purely light-driven charge pumping of hot electrons and its THz radiation [Fig.~\ref{fig:fig5}(a)], but it becomes {\em crucial} for enhanced THz radiation in Figs.~\ref{fig:fig5}(b) and ~\ref{fig:fig5}(c) when LLG dynamics of LMMs within Mn$_3$Sn is turned on. This suggests that {\em charge pumping by the dynamics of LMMs in the presence of its intrinsic SOC}~\cite{Mahfouzi2012,Ciccarelli2015}, rather than standardly assumed spin pumping plus spin-to-charge conversion away from magnetic layer~\cite{Rouzegar2022,Seifert2022,Bull2021,Wu2021,Seifert2016,Wu2016,Chen2018,Nenno2018,Nenno2019}, can be an important and largely unexplored resource for spintronic THz emitters. This resource is offered by the class of materials (such as Mn$_3$Sn or two-dimensional magnetic materials~\cite{Gibertini2019})  possessing   
	strong SOC in their bulk. 
	
	Although not in the focus of detection schemes in THz spintronics, we also examine outgoing EM radiation in the frequency range of the incoming laser pulse. The bottom row of panels in Fig.~\ref{fig:fig5} reveals {\em both} even and odd high harmonics in the FFT power spectrum up to order $n \le 7$. The width of their peaks is broadened upon turning LLG dynamics on. Let us recall that  high-harmonic generation has been intensely pursued in recent years for solids driven out of equilibrium by laser light of frequency $\Omega_0$~\cite{Ghimire2019,Tancogne-Dejean2022,Yamada2021}. For example, inversion symmetric bulk semiconductors driven by strong mid-infrared laser light, whose $\hbar \Omega_0$ is much smaller than the band gap, can exhibit nonlinear effect generating new EM radiation at odd multiplies of $\Omega_0$~\cite{Ghimire2019,Tancogne-Dejean2022,Yamada2021}. Furthermore, in two-dimensional 2D systems breaking inversion symmetry, such as ML of MoS$_2$~\cite{Ghimire2019} or surface states of topological insulators~\cite{Bai2021}, additional even-order harmonics or  non-integer harmonics~\cite{Bai2021} can emerge. The Weyl electrons of Mn$_3$Sn driven by light of frequency $\Omega_0$ are predicted in Fig.~\ref{fig:fig5} to radiate  {\em both} even and odd harmonics at frequencies $\Omega=n\Omega_0$, because two MLs of Mn$_3$Sn used in Fig.~\ref{fig:fig1} break inversion symmetry.  It was also conjectured theoretically~\cite{Jia2020} that high harmonic spectrum can be used to probe underlying magnetic configuration-mediated topological phases of Weyl semimetals, which is yet to be explored for Mn$_3$Sn. The important role played by the lattice symmetries of Mn$_3$Sn and noncollinear configuration of their LMMs for high-harmonic generation is confirmed by high-harmonic spectrum being largely insensitive to switching off SOC within Mn$_3$Sn layer [light gray curves in Figs.~\ref{fig:fig5}(e) and ~\ref{fig:fig5}(f)].

   \subsection{Emitted EM radiation: Poynting vector and efficiency} \label{sec:poynting}
	
    In order to quantify efficiency of conversion of the power of incoming radiation into the power of outgoing radiation, we use $\mathbf{E}$ and $\mathbf{B}$ fields from the Jefimenko equations to compute the Poynting vector~\cite{Poynting1884} measuring the energy flux (i.e., the energy transfer per unit area per unit time) of an electromagnetic field. Although the concept of the Poynting vector has been known for well over a century~\cite{Poynting1884}, the actual flow of electromagnetic
    energy in even the simplest circuits is a subject of ongoing interest~\cite{Ridley2021,Galili2005,Harbola2010}. For much more complicated case of laser-driven magnetic bilayers, calculations of the Poynting vector of incoming and outgoing electromagnetic radiation and thereby defined efficiency of their conversion into each other, is lacking in THz spintronics literature.

\begin{figure}
		\centering
		\includegraphics[width=\linewidth]{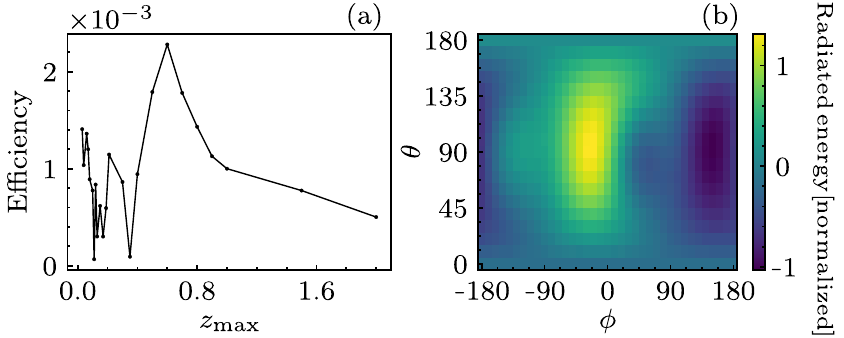}
		\caption{(a) The output-to-input radiation conversion efficiency  from bilayer in Fig.~\ref{fig:fig1}, defined as the ratio of outgoing to incoming total radiation energy within relevant time interval $t_i=6$~ps to $t_f=10$~ps, as a function of the $z_{\rm max}$ [Eq.~\eqref{eq:peierls}] measuring the intensity of the input laser pulse. (b) The angular dependence of (normalized) total  radiated energy into the  solid angle $d\Omega$ at a distance 100$a_0$ for $z_{\rm max}=0.12$. The abscissa  and the ordinate axes are  azimuthal $\phi$ and polar $\theta$ angle, respectively.}
		\label{fig:fig7}
	\end{figure}

	The time-dependent energy flux from the computed electric $\mathbf{E}(\mathbf{r},t)$ and magnetic  $\mathbf{B}(\mathbf{r},t)$ vector fields is given by the Poynting vector
	\begin{equation}\label{eq:poynting}
		\mathbf{S}(\mathbf{r},t) = \frac{1}{\mu_0}\mathbf{E}(\mathbf{r},t) \times \mathbf{B}(\mathbf{r},t).
	\end{equation}
	From it, we can obtain the power $dP =\mathbf{S}\cdot \hat{\mathbf{R}}R^2d\Omega$ radiated into the solid angle $d\Omega=\sin \theta d\theta d\phi$ at  distance \mbox{R=100 $a_0$} in the direction specified by the unit vector $\hat{\mathbf{R}}$. This quantity, integrated over the time interval within which radiation exists in Figs.~\ref{fig:fig3} and ~\ref{fig:fig4} gives the output energy $E_{\rm out}^{\theta,\phi} = \int_{t_i}^{t_f}\!dt dP$ plotted in Fig.~\ref{fig:fig7}(b) as a function of the azimuthal $\phi$ and polar  $\theta$ angles, after normalization $[E_{\rm out}^{\theta,\phi}]_\mathrm{norm} = (E_{\rm out}^{\theta,\phi} - [E_{\rm out}^{\theta,\phi}]_\mathrm{mean})/(2\sigma_{E_{\rm out}^{\theta,\phi}})$ 
	where $[E_{\rm out}^{\theta,\phi}]_\mathrm{mean}$ is the average value of $E_{\rm out}^{\theta,\phi}$ within the interval $[t_i,t_f]$ of discrete time points and $\sigma_{E_{\rm out}^{\theta,\phi}}$ is its standard deviation.  To get the total radiated output energy, we integrate $dP$ additionally over the whole space $E_{\rm out} = \int_{t_i}^{t_f}\!dt \int_{\theta,\phi} \!dP$, where \mbox{$t_i$=6 ps} to \mbox{$t_f$=10 ps}. The total input energy $E_{\rm in}$ is the integral of $dP$ of the input laser pulse over the area of the Mn$_3$Sn surface and over the same time interval. Then the output-to-input efficiency is defined by $|E_{\rm out}|/|E_{\rm in}|$ which is plotted in Fig.~\ref{fig:fig7}(a) as the function of $z_{\rm max}$ [Eq.~\eqref{eq:peierls}] measuring the intensity of the input laser pulse. Figure~\ref{fig:fig7}(a) shows that the efficiency of conversion is rather small, but it can be tuned by a factor of two by changing $z_{\rm max}$.
	
	 %----------------------------------------------------------
	%----------------------------------------------------------
	
	\section{Conclusions}\label{sec:conclusions}
	
	In conclusion, we have developed a fully quantum (based on the Lindblad QME) and fully microscopic (based on an input single-particle or many-body  Hamiltonian) treatment of electrons within magnetic bilayers in THz spintronics, which is combined in multiscale fashion with classical LLG treatment 
	of LMMs and classical Maxwell treatment of incoming and outgoing EM radiation, as illustrated by Fig.~\ref{fig:fig2}.  Despite enormous experimental and technological interest in spintronic THz emitters~\cite{Rouzegar2022,Seifert2022,Bull2021,Wu2021,Seifert2016}, as well as fundamental and theoretical interest in far-from-equilibrium~\cite{Gillmeister2020,Wang2017a} magnetically ordered quantum materials driven by ultrafast light, very few theoretical approaches 
	have demonstrated complete treatment of phenomena on different time and length scales in these systems, that is,  by  starting  from input fs laser pulse and 
	by proceeding to compute the output THz radiation. For example, we are aware of one such complete treatment~\cite{Nenno2019} where semiclassical transport theory (based on the Boltzmann equation) for incoming laser pulse-driven hot electrons is combined with the Maxwell equations for outgoing EM radiation. In that approach, hot electrons carry only spin current which is assumed to be converted into charge current (by the inverse spin Hall effect) radiating in the THz range~\cite{Nenno2019}. However, this approach neglects pumping of charge currents of hot electrons directly by light~\cite{Bajpai2019} and prior to any spin-to-charge conversion, as well as that spin current can have different contributions (as observed in very recent experiments~\cite{Jimenez-Cavero2022}) many of which cannot~\cite{Tserkovnyak2005,Chen2009} be captured by semiclassical transport theory.

	Importantly, spin~\cite{Tserkovnyak2005,Chen2009} and charge~\cite{Mahfouzi2012,Chen2009} currents pumped by the dynamics of classical 
	LMMs surrounded by conduction electrons require fully quantum transport treatment, as provided by standard approaches like the scattering theory~\cite{Tserkovnyak2005} or nonequilibrium  Green functions~\cite{Chen2009,Petrovic2018,Petrovic2021,Bajpai2020}, with additional effort required~\cite{Mahfouzi2012,Dolui2020a,Dolui2022,Chen2015}  when SOC is present within the magnetic layer or at its interfaces~\cite{Tserkovnyak2005} with other materials. As demonstrated by Figs.~\ref{fig:fig3} and ~\ref{fig:fig4}, QME+LLG+Jefimenko approach developed in this study captures spin and charge pumping due to laser pulse, as well as due to subsequent much slower LMM dynamics. By switching on and off different terms in the quantum and classical Hamiltonians, as the input of  QME+LLG+Jefimenko  approach, we reveal the following features in ultrafast light-driven Weyl AF Mn$_3$Sn in contact with nonmagnetic layer: ({\em i}) both spin and charge currents are initially pumped by the laser pulse, where spin current subsequently exerts spin torque on slow (movie in the SM~\cite{sm}) LMMs, which in turn pump additional spin and charge currents enhancing [Figs.~\ref{fig:fig3} and ~\ref{fig:fig4}] those initially driven by light; ({\em ii}) consequently, additional charge current pumping by LMM dynamics enhances EM radiation in THz range [Fig.~\ref{fig:fig5}(b)], while broadening light-driven high harmonics [Fig.~\ref{fig:fig5}(e)] of the pulse center frequency; ({\em iii}) for the specific magnetic bilayer in Fig.~\ref{fig:fig1}, the {\em key contribution} to generated charge current and its THz radiation comes from {\em charge pumping by LMMs in the presence of intrinsic SOC} of Mn$_3$Sn  [Fig.~\ref{fig:fig5}(b)], rather than from conversion~\cite{Jungfleisch2018a} of pumped spin current into charge current by interfacial SOC which can, in fact, even be detrimental to emitted THz radiation [Fig.~\ref{fig:fig5}(b) vs. \ref{fig:fig5}(c)].
	
	Although a single mechanism cannot explain vastly different~\cite{Koopmans2010,Tauchert2022} magnetic bilayers, we believe that some of our conclusions for  bilayers of Mn$_3$Sn apply also to other systems whose examination via QME+LLG+Jefimenko approach we relegate to future studies. We also relegate to future studies  magnetic bilayers where localized spins must be treated quantum-mechanically and only one input---Hamiltonian~\cite{Tows2019,Bajpai2021} of electrons interacting with quantum localized spins---would be required for the QME+Jefimenko calculations. 
	
	%======================================ACKNOWLEDGEMENTS==============================
	\begin{acknowledgments}
		This research was primarily supported by NSF through the University of Delaware Materials Research Science and Engineering Center, DMR-2011824.
	\end{acknowledgments}
	
	%==========================================================
	%********************references***************************
	
	%\bibliography{qttg}
	
	%\bibliographystyle{C:/BIBTEX/IEEEtran}
	%\bibliography{C:/BIBTEX/qttg}

\end{document}